\documentstyle[11pt,aaspp4,flushrt]{article}

\def\Fe{Fe K$\alpha$ }
\def\g{_{\rm g}}
\def\pc{{\rm pc}}
\def\kms{{\rm km/s}}

\def\ltsima{$\; \buildrel < \over \sim \;$}
\def\simlt{\lower.5ex\hbox{\ltsima}}
\def\gtsima{$\; \buildrel > \over \sim \;$}
\def\simgt{\lower.5ex\hbox{\gtsima}}

\received{}
\revised{}
\accepted{}
\cpright{type}{}

\lefthead{}
\righthead{}

\begin{document}

\title{Fe K$\alpha$ line: A tool to probe massive binary black holes
in Active Galactic Nuclei?}
\author{Qingjuan Yu}
\affil{Princeton University Observatory, Princeton, NJ 08544-1001, USA\\ Email:
yqj@astro.princeton.edu}
\and
\author{Youjun Lu}
\affil{Center for Astrophysics, Univ. of Sci. \& Tech. of China, Hefei,
Anhui 230026, P.R. China\\ Email: lyj@astro.princeton.edu}

\begin{abstract}

Hierarchical mergers of galaxies can form binary black holes (BBHs) since
many or most galaxies have central massive black holes (BHs).
It is possible that some BBHs exist in active galactic nuclei (AGNs).
We argue that each BH may be surrounded by an accretion disc with a different
inclination angle to the line of sight (due to different BH spin directions
and the Bardeen-Petterson effect).
The observed \Fe line profile from a BBH system is a combination of the lines
from the inner regions of the two discs, which is significantly affected by
the inclination angles of the two discs.
The \Fe line profile associated with BBHs may have an unusual shape with double
or more peaks as well as short-term variability, which can be distinguished
from the \Fe line properties of some other possible models.
We suggest that with the improvement of resolution in X-ray astronomy, \Fe line
profiles be a potential tool to probe the existence of massive BBHs in AGNs.
The \Fe line profile associated with BBHs may also provide a tool to
investigate the dynamics in strong gravitation field
(e.g. providing evidence of the Bardeen-Petterson effect).
\end{abstract}

\keywords{
Black hole physics--Accretion--Line profile--galaxies: active
}

\section{Introduction}

\noindent
Much evidence indicates that massive black holes (BHs) reside in the centers
of many or most galaxies (e.g. \cite{Metal98}). Mergers of galaxies are
likely to form massive binary black holes (BBHs). Theoretical estimation
shows that the BBH lifetime is not much shorter than the Hubble
time and many BBHs should be still in the centers of galaxies (\cite{BBR80}).
The existence of BBHs in the universe will not only provide a laboratory to
test gravitation radiation theory and BH physics, but also
probe of the hierarchical structure model of galaxy and large-scale
structure formation.

Currently, there is no systematic and unambiguous method to identify BBHs.
BBHs stay at a separation in the range $10^{16}-10^{19}$cm (e.g. $10^{-4}
-10^{-1}$ arcsec at 10Mpc) during the slowest evolution period 
(\cite{BBR80,Y02}), 
and thus it is hard to resolve a BBH --- two very close galactic nuclei in
the image --- with current telescope resolution.
The shallow cusps in the inner surface brightness profiles of some nearby giant
galaxies may be produced from steep cusps by ejecting stars from their inner
regions during the hardening of BBHs (\cite{Fetal97,QH97}), but there is still
no proof of a currently existing BBH in those galaxies.
To identify a BBH, we have to find some other
effect of BBHs on their nearby environment and/or some
manifestation of the motion in a two-body system, such as jet precession
(\cite{BBR80}), double-peaked Balmer lines (\cite{GASK96}), or
quasi-periodic radio, optical, X-ray or $\gamma$-ray variation (e.g. OJ287:
Sillanp\"{a}\"{a} et al. 1988, Valtaoja et al. 2000; Mkn501: 
Rieger \& Mannheim 2000).
Some active galactic nuclei (AGNs) have been claimed to be detected as BBH
candidates by those methods, but all of them are controversial because of
other explanations for the same phenomena or some inconsistency with other
observational evidence.

If there is sufficient gas with some angular momentum close to a BBH, we may
expect that the gas in the vicinity of each BH is accreting onto
the BH in the form of a disc rather than in the form of spherical accretion,
which may make the system appear as an AGN. If the binary separation is small
(say, much less than the scale of the broad line region),
the two accretion discs may be warped at outer parts and connected with an
outer large circumbinary accretion disc. If the separation is large enough,
each BH is probably accompanied by its own disc and broad line region. The spin
axes of the two BHs are very likely to be misaligned (\cite{R78,BBR80}) 
and the discs in this two-accretion-disc (TAD) system
associated with the BBH can also have different inclination angles to the
line of sight.

AGNs are observed to be copious X-ray emitters. This X-ray emission is
believed to originate from the very inner accretion disc around a
massive BH. The broad skewed iron K$\alpha$ line profile found by {\it GINGA}
and confirmed by {\it ASCA} is believed to result from a combination
of gravitational broadening and Doppler shift in an accretion disc
(\cite{TAN95}). So far, alternative models have failed to account for this
profile (\cite{FAB95,FAB00}), which offers one of the strongest lines of
evidence for the existence of massive BHs. X-ray spectroscopy also
promises a powerful tool to detect strong-gravitation-field relativistic
effects in the vicinity of a massive BH. The observed \Fe line profile
is significantly affected by the inclination angle of the disc to the
observers line of sight (\cite{FAB89,L91}). In a BBH system, the observed
\Fe line profile can be a combination of the line profiles from two discs
with very different inclination angles to the line of sight. Motivated by this
observation, we propose a method to probe BBHs in AGNs by \Fe line
profiles. We expect that \Fe line profiles will become an efficient way to
probe BBHs in AGNs.

\section{Two accretion discs in BBH systems}

\noindent
\begin{figure}
 \epsscale{0.6}
 \plotone{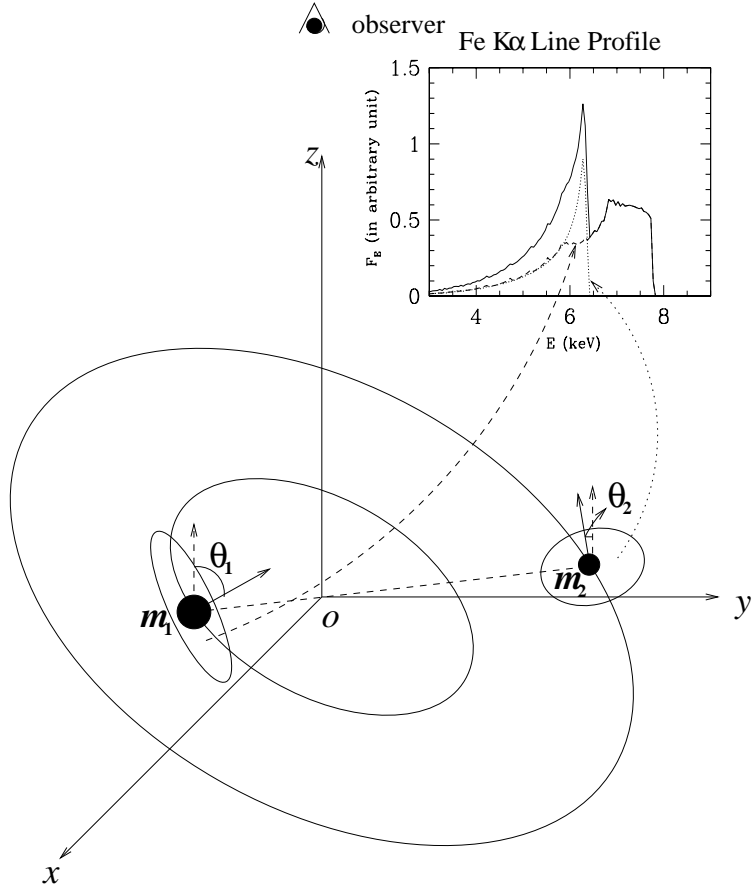}
 \caption{Schematic diagram of a BBH and its accretion discs: the BBH
containing BHs $m_1$ and $m_2$ is rotating around the center of mass `O'
in circular orbits. A distant observer is located in the $z$-axis direction.
The inclination angles of the two discs to the line of sight are $\theta_1$ and
$\theta_2$, respectively.
An example of \Fe line profile is plotted in the upper-right of this figure
by setting $\theta_1=60^{\rm o}$, $\theta_2=5^{\rm o}$, the emissivity ratio
$\epsilon^0_2/\epsilon^0_1=0.6$ (or equivalently the mass ratio $m_2/m_1=0.6$
by assuming the two discs have almost the same dimensionless accretion
rates), and $p=2.5$ (the exponent of the emissivity law in \S 3).
The dashed line represents the component from the disc of the BH $m_1$ and the
dotted line represents the component from the disc of the BH $m_2$.
The solid line is the observed line profile, which is a combination of
the two components.
}
 \label{fig:geom}
\end{figure}

\noindent
Consider a BBH containing BHs of mass $m_1$ and $m_2$ rotating around their
center of mass (Fig.~1), the relative orbit of the two BHs is assumed to be
circular with separation $a$. The Keplerian orbital period of the binary is
\begin{equation}
P^{\rm orb} =  210\left(\frac{a}{0.1\pc}\right)^{3/2}
\left(\frac{2\times 10^8M_\odot}{m_1+m_2}\right)^{1/2} {\rm yr}
\end{equation}
and their maximum orbital velocities relative to the center of mass in the
line of sight are
\begin{equation}
|v_i|=1.5\times10^3\kms\left(\frac{0.1\pc}{a}\right)^{1/2}\left(\frac{m_1
+m_2}{2\times10^8M_\odot}\right)^{1/2}\left[\frac{2m_1m_2}{m_i(m_1+m_2)
}\right]\sin \theta_{\rm orb} \ \ \ \ (i=1,2)
\label{eq:vel}
\end{equation}
where $\theta_{\rm orb}$ is the orbital inclination to the line of sight.
A moderate orbital eccentricity of the BBH orbit will not change the above
values much. The orbital motion of each BH is ignored in the later
calculation of \Fe line profiles since it only causes a slight overall
Doppler shift in the profiles.

For a single rapidly rotating BH with mass $m$, the Bardeen-Petterson effect
(\cite{BP75}) can cause the gradual transition of the disc into the BH 
equatorial plane in the region between radii $10^4r\g$ and $10^2r\g$ 
(note that there are still controversies on the location of the transition
radii, c.f. Nelson \& Papaloizou 2000 and references therein), where
$r\g=Gm/c^2$. Thus a planar disc could be formed in the equatorial
plane of the BH. The timescale for the BH to align with the outer disc
angular momentum cannot be estimated precisely, but is of the order
of the overall lifetime of AGNs (Rees 1978, Nelson \& Papaloizou 2000; 
however, a much smaller realignment timescale of the order of $10^5$ yr is
gotten by Natarajan \& Pringle 1998). In a BBH system, it is likely that both
BHs have large spins and the spin axes can be in very different directions,
especially in the early phase of the nuclear activity. This difference may
not be changed in a time much shorter than the lifetime of the nuclear
activity. Therefore, the two accretion discs associated with a BBH can
have different inclination angles to the line of sight. Even if only one
of the BHs has a large spin, it is still very likely that its spin axis
is different from the disc direction of the other (Schwarzschild) BH.

The Roche radius of the smaller BH is about $(m_2/3m_1)^{1/3}a$ if $m_1>m_2$.
If the two BHs have comparable masses and a typical separation of
$\sim 10^4G(m_1+m_2)/c^2$, the inner region of each disc, at least within 
$r\simlt10^3 Gm_i/c^2$,
is not easily disrupted by tidal forces from the other BH. In a BBH system,
the spin axis of each BH will undergo geodesic precession about its total
angular momentum as discussed by Begelman, Blandford \& Rees (1980). When
the spin angular momentum of the binary is small compared with its orbital
angular momentum ($m_2/m_1\simgt\sqrt{Gm_1/ac^2}$ if $m_1>m_2$),
the spins will precess in a cone with the orbital angular momentum as axis.
Their precession periods are given by:
\begin{equation}
P^{\rm prec}_i\sim 6\times 10^6\left(\frac{a}{0.1{\rm pc}}\right)^{5/2}
\left(\frac{m_1+m_2}{2\times10^8 M_\odot}\right)^{1/2}
\left(\frac{10^8 M_\odot m_i}{m_1m_2}\right)^2{\rm yr} \ \ \ \ (i=1,2).
\end{equation}
The precession may cause the two accretion discs to be tilted to their BHs
equatorial plane, but the two disc inclination angles to our line
of sight will not change in a short time.
Detailed study of the dynamics and stability of the TADs in BBHs is beyond
the scope of this paper.

\section{Emergent Fe K$\alpha$ line profiles from BBH systems}

\noindent
We have argued that there are good reasons for the existence of TADs with
different inclination angles to the line of sight in BBHs. We shall assume
that both discs are cold thin accretion discs. The observed \Fe line profile
is then the summation of the two components from the TADs. The combined line
profile is mainly controlled by the inclination angles of the TADs and the
relative strength of the two components. The relative strength is related to
the mass ratio of the two BHs and the accretion rates onto them (if
the two accretion systems have almost the same dimensionless accretion rate
$\dot{m}=\dot{M}/\dot{M}_{\rm Edd}$,
where $\dot{M}$ is the accretion rate and
$\dot{M}_{\rm Edd}$
is Eddington accretion rate,
the relative strength of the two components will be given by the mass ratio
$m_1/m_2$). 

Using the ray-tracing technique and elliptic integrals (\cite{RB94}), we
follow the photons from each accretion disc to a distant observer, and
calculate the corresponding redshift of the photons and the resulting line
profiles. The spins of BHs are both set to be $a/M=0.998$.
The \Fe line photons are assumed to be isotropically emitted in the frame
moving with the accretion disc material, and the surface emissivity is given
by the power-law $\epsilon_i(r)=\epsilon_i^0r^{-p}$ for the region 
$r^{\rm in}_i\le r\le r^{\rm out}_i$. We adopt the averaged line emissivity
exponent index of $p=2.5$ which is obtained from the fitting of \Fe
line profiles for a sample of AGNs observed by {\it ASCA} (Nandra et al. 1997).
We set the inner radius to be the marginally stable orbit $r_{\rm ms}$
(few \Fe line photons come from the region inside the marginally stable orbit
for Kerr BH with spin 0.998), and the outer radius $r_i^{\rm out}=
160Gm_i/c^2$ (the line profiles are not sensitive to the value of
$r_i^{\rm out}$ since most line photons come from the inner region for
a typical line emissivity law with $p=2.5$).
An example of the resulting spectral line is shown in Figure~1, for discs with
inclination angles 
$\theta_1=60^{\rm o}$
and
$\theta_2=5^{\rm o}$,
and
$\epsilon_2^0/\epsilon_1^0\sim m_2/m_1\sim 0.6$.
This unusual line profile is double-peaked, asymmetric and has two `edge'-like 
feature. The peak with the
smaller central energy, near the rest frame energy of \Fe line -- 6.4keV,
(or less than 6.4keV due to gravitational redshift for extremely low
inclination disc), comes from disc D$_2$ with a small inclination angle, has
the characteristic features of emission from a relativistic accretion disc:
a skewed red wing and a sharp ``blue'' edge (note that the energy of the 
``blue'' edge is mainly determined by the inclination of a disc, and the
red wing extent is sensitive to the inner radius of the line emission region).
The peak with the smaller central energy is
narrower than the broad component, which comes from disc D$_1$ with a
high inclination angle. If the spin of the BH $m_1$ is smaller, the ``narrow
component'' can be even narrower. The ``narrow'' component can be stronger
or weaker than the broad component depending on the relative emissivity.
Another important feature is that these two components should both have {\it
short-term variability} of intensity and shapes on the timescale of $10^4$s,
as suggested by the variation in some Seyfert galaxies (\cite{IWA96,NAD99}),
and their variation patterns can be totally different and unrelated with
each other since they come from two different discs. All of the above
features make it easy to distinguish the BBH model from the other
alternative models producing a two-component line shape. 

\begin{figure}
 \epsscale{0.8}
 \plotone{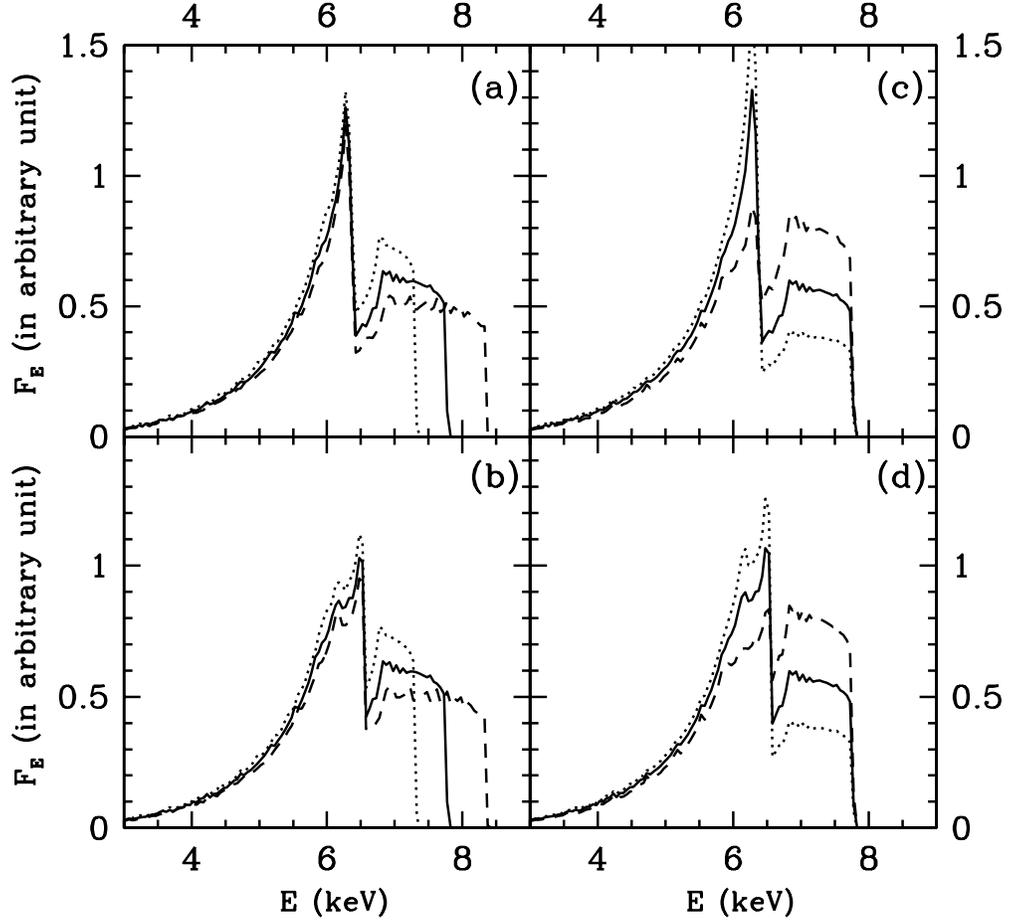}
 \caption{\Fe line combined profiles from TADs in BBHs.
The exponent index of the emissivity law $p$ is 2.5.
In panels (a) and (b) $\epsilon^0_2/\epsilon^0_1=0.6$,
$\theta_1=50^{\rm o}$ (dotted line), $60^{\rm o}$ (solid line),
$70^{\rm o}$ (dashed line), but $\theta_2=5^{\rm o}$ in (a) and
$\theta_2=20^{\rm o}$ in (b).
In panels (c) and (d) $\epsilon^0_2/\epsilon^0_1=1.5$ (dotted line), 0.7
(solid line), 0.2 (dashed line), $\theta_1=60^{\rm o}$, but $\theta_2=5^{\rm o}$
in (c) and $\theta_2=20^{\rm o}$ in (d).
}
\label{fig:line}
\end{figure}

Not all of the emergent \Fe line profiles from TAD systems are so
distinguishable from the profiles produced by only one disc. If the relative
emissivity of the two discs is too small or too large, the component emitted
from one accretion disc in TADs system will be drowned by the other. Only
when 
$\epsilon^0_2/\epsilon^0_1$
is about in the range $0.2-2$ (or $m_2/m_1$ is in a similar range if
luminosity is proportional to Eddington luminosity) will the line
profiles from TADs be significantly different from the one emitted by a
single BH--accretion disc system. If $\epsilon^0_2/\epsilon^0_1$
is a little less than $0.2$, the line profile may be misunderstood as a
relativistic line from a single disc plus an absorption feature; and if
$\epsilon^0_2/\epsilon^0_1$ is a little larger than $2$, the line profile
may be misunderstood as a relativistic line from a single disc plus
a high ionization line (e.g. Fe K$\beta$ or Ni K$\alpha$ line).
If the difference between
$\theta_1$ and $\theta_2$ is too small, the two line components will be
blended, and thus difficult to distinguish from the profile produced by a
single BH--accretion disc system. The line profiles in TAD systems
can be very complicated with double peaks, three peaks and even four peaks
depending on the inclination angles of the two accretion discs and the
relative emissivity (Fig. 2). When
$\theta_1\sim45^{\rm o}-70^{\rm o}$
and
$\theta_2\sim0^{\rm o}-20^{\rm o}$,
the emergent line profile will clearly exhibit two distinct components
(see Fig. 2; note that the line from a disc with very high inclination
angle (say, $>70^{\rm o}$) may be strongly affected by the limb-darkening
effect of the outer layer of accretion disc). If the inner disc is somewhat
ionized, or the emissivity law is somewhat different, the line profile
should remain similar.
The probability that the combined line from two randomly oriented equal discs
is in the shape similar to those shown in Figure~2, is about 20\%
(the difference between the inclination angles is larger than $30^{\rm o}$
and both the disc inclination angles are not larger than $70^{\rm o}$).
The amounts of AGN sources expected to harbor BBH systems with comparable
BH masses, which could be identified by \Fe line profiles, are relevant 
to the process of structure formation and the merger history of galaxies.

The line shapes from the TAD systems shown in Figure~1 and Figure~2 are examples
chosen from many idealizations. We have neglected such complication as
the actual dynamics of the system, the possible warp of the outer disc, the
ionization of the accretion discs, the real line emissivity law, contamination
from non-disc emitters, the irradiation of one disc by the X-ray photons from
the other, and the absorption of intervening gas and dust etc.
The complication certainly affects the line profile quantitatively, 
but will not make much difference qualitatively. 

\section{Differentiating the BBH model from other possible models}

\noindent
There are other possible models that can produce two-component line profiles,
but they are not difficult to differentiate from the BBH model.

First, off-axis X-ray flares above a single disc can strongly affect the line
profiles (off-axis-flare model; c.f. Yu \& Lu 2000).
In \S 3, the disc emissivity is assumed to be axisymmetric.
In reality, X-ray flares can be off the disc rotation axis and the emissivity
law of the disc is non-axisymmetric because the disc region just under the 
flares generally receive more illumination.
Thus, the non-axisymmetric emissivity may strongly affect the \Fe line profiles 
since different energy parts of \Fe line profiles stem from
different regions of a disc.
For example, a cold accretion disc is illuminated by two local flares: one is
atop the approaching side of the disc, where the energy of \Fe line photons
are blue shifted by Doppler effect; and the other is atop the disc region
towards or backwards us which corresponds to the line emission around 6.4keV.
This situation can make line shapes like the one from TADs.
The X-ray flares are probably produced
by some thermal instability in the disc or magnetic reconnection and their
locations should be randomly distributed. The appearance of the line profile
arising from flares can be very complex. However, the average line profile
over several flares should be consistent with the profile from a single disc
(with a single inclination angle),
as suggested by the observations of MCG-6-30-15 (\cite{IWA96,IWA99}),
which is quite different from the combined line profiles from TADs.

Second, a two-component line shape can be produced if a narrow line
component is emitted from a molecular torus or broad line region (BLR) clouds,
and a broadened line component is emitted from the inner region of a
highly inclined accretion disc (BLR/Torus-AD model, c.f. \cite{Y01}).
In this scenario, the central energy of the narrow component should be
around 6.4keV or a little blue-shifted by the outflow motion, and the line
width caused by the velocity
dispersion of the clouds should be several thousand \kms.\footnote{In AGNs,
the blueshift of high ionization lines (e.g. CIV line ) suggests that the
outflow of the line emission gas in BLR has a velocity from zero to several
thousand \kms and no evidence about the inflow motion of emission materials
in BLR has been found (\cite{SUL00}).}
In contrast, in the BBH model, the ``narrow'' component is redshifted to 
energies less than 6.4keV if the inclination is very low (due to gravitational
redshift). With higher inclination, the narrow component is
centered around 6.4keV or higher, but with a large width. More importantly,
in the BLR/Torus-AD model, the narrow component should remain constant on the
timescale of several days or more (the light-travel time across the BLR or
the torus) rather than varying on timescales $<10^4$s, as predicted by
the BBH model. 

Third, a two-component line shape can also be produced by the combination of
a broadened iron line from a highly inclined accretion disc plus a component
which is scattered into our line of sight by an efficient electron-scattering
material (Scatter-AD model, c.f. Wang et al. 1999). It is unlikely that
this electron-scattering material is close to the disc because this
configuration is unlikely to produce a line with two distinct components:
first, the scattered line component shape will be different from the one
observed from a disk with a small inclination angle; second, \Fe photons
from the inner accretion disc cannot be seen directly because the disc is
covered by the scattering material. 
If the scattering layer is high enough above the disc, the scattered line
profile would be somewhat like the one observed at a small inclination
angle; however, the covering factor would normally be small and the
equivalent width of this component should therefore be much smaller than
the broad one since the scattered photons are redistributed in all
directions. In contrast, in the BBH model, the equivalent width of the
``narrow'' component can be larger than that of the broad component.
Furthermore, the temporal variations of the two components in the Scatter-AD
model must be strongly correlated with a time delay which reflects the
distance between the scattering material and the disc.

\section{Prospects}

\noindent
To use \Fe profiles to identify a BBH with TADs, we need to resolve 
both the narrow component and broad component of the line profiles and study
the variation of the line profiles and intensity with time. Over the past
decade, a number of Seyfert galaxies and QSOs have been shown to have broad
\Fe lines (\cite{NAD97,YQO00}). The \Fe line profile of NGC4151 is fitted by
two disc components with inclinations of $0^{\rm o}$ and $58^{\rm o}$
better than by a one-disc line model (\cite{WANG99}). The component with
inclination $0^{\rm o}$ was explained by being scattered into our line of
sight; however, those two components can also be explained as coming from
possible TADs. Future observations with higher resolution are needed to check
if NGC4151 is a possible BBH candidate. The unique \Fe profile of MARK205
(see Fig. 2 in Reeves et al. 2000), recently revealed by {\it XMM-Newton}, is
somewhat like the line shape established in Figure~1. Its broad line component
can be fitted by an accretion-disc line with a high inclination. Unfortunately,
the narrow component is not fully resolved. The present observations
can be explained by assuming that the narrow component comes from neutral
matter at large distant from a central BH and the broad one is emitted
from a highly ionized relativistic accretion disc (\cite{REEV00}). So, further
observations and variability studies are needed to check if this object is
a BBH candidate or not. 

Stimulated by those special line shapes observed in some objects, we believe
that some BBHs in AGNs, if any, can be identified by searching the unusual
iron line profile with current and future X-ray satellites, such as 
{\it XMM-Newton, Constellation-X and XEUS}. 
If any one of AGNs is revealed to have the typical
line shapes shown in Figure~2 as well as short-term variability, it will
provide one of the strongest lines of evidence for the existence of BBHs
and a laboratory to investigate the dynamics in strong gravitation field
(e.g. the Bardeen-Petterson effect).

\acknowledgments{
We are grateful to Scott Tremaine for a careful reading
of an early draft with many insightful comments and suggestions.
We thank Jeremy Goodman and David N. Spergel for helpful discussions.
YL acknowledges the hospitality of the Department of Astrophysical Sciences,
Princeton University.
}

{}

\end{document}